\documentclass[aps,prx,floatfix,superscriptaddress,twocolumn,footinbib]{revtex4-1}
\usepackage{subfigure}
\usepackage{amssymb}
\usepackage{amsmath}
\usepackage{amsfonts}
\usepackage{appendix}
\usepackage{bm}
\usepackage{graphicx}
\usepackage{epsfig}
\usepackage{epstopdf}
\usepackage{balance}
\usepackage[dvipsnames]{xcolor}
\usepackage{calc}
\usepackage{natbib}
\usepackage{color}
\usepackage[colorlinks,
            linkcolor=blue,
            anchorcolor=blue,
            citecolor=blue,
            urlcolor=blue]{hyperref}

\usepackage{lipsum}

\begin{document}

\title{Orbital-FFLO State and Josephson Vortex Lattice Melting in Layered Ising Superconductors}

\author{Hongyi Yan}
\affiliation{Center for Advanced Quantum studies, Department of Physics, Beijing Normal University, Beijing 100875, China}
\affiliation{Key Laboratory of Multiscale Spin Physics, Ministry of Education, Beijing,
100875, China}
\author{Haiwen Liu}
\email{haiwen.liu@bnu.edu.cn}
\affiliation{Center for Advanced Quantum studies, Department of Physics, Beijing Normal University, Beijing 100875, China}
\affiliation{Key Laboratory of Multiscale Spin Physics, Ministry of Education, Beijing,
100875, China}
\affiliation{Interdisciplinary Center for Theoretical Physics and Information Sciences, Fudan University, Shanghai 200433, China}
\author{Yi Liu}
\affiliation{Department of Physics and Beijing Key Laboratory of Opto-electronic Functional Materials \& Micronano Devices, Renmin University of China; Beijing 100872, China}
\affiliation{Key Laboratory of Quantum State Construction and Manipulation (Ministry of Education), Renmin University of China; Beijing 100872, China}
\author{Ding Zhang}
\affiliation{State Key Laboratory of Low Dimensional Quantum Physics and Department of Physics, Tsinghua University, Beijing 100084, China}
\affiliation{Beijing Academy of Quantum Information Sciences, Beijing 100193, China}
\affiliation{Frontier Science Center for Quantum Information, Beijing 100084, China}
\affiliation{RIKEN Center for Emergent Matter Science (CEMS), Wako, Saitama 351-0198, Japan}
\author{X. C. Xie}
\affiliation{Interdisciplinary Center for Theoretical Physics and Information Sciences, Fudan University, Shanghai 200433, China}
\affiliation{International Center for Quantum Materials, School of Physics, Peking University, Beijing, China}
\affiliation{Hefei National Laboratory, Hefei, China}

\date{\today}

\begin{abstract}
This study explores the impact of in-plane magnetic fields on the superconducting state in layered Ising superconductors, resulting in the emergence of the orbital Fulde-Ferrell-Larkin-Ovchinnikov (FFLO) state coupled with Josephson vortices. Recent experiments have revealed an unexpected first-order phase transition in these superconductors under strong in-plane magnetic fields. Our theoretical analysis demonstrates that this phase transition is primarily driven by the formation and subsequent melting of a Josephson vortex lattice within the superconducting layers. As the magnetic field increases, the vortex lattice undergoes a transition from a solid to a liquid state, triggering the observed first-order phase transition. We calculate both the melting line and the in-plane critical field in the phase diagram, showing strong agreement with experimental results.
\end{abstract}
\maketitle

\begin{figure*}
    \centering
    \includegraphics[width=15cm,height=8.5cm]{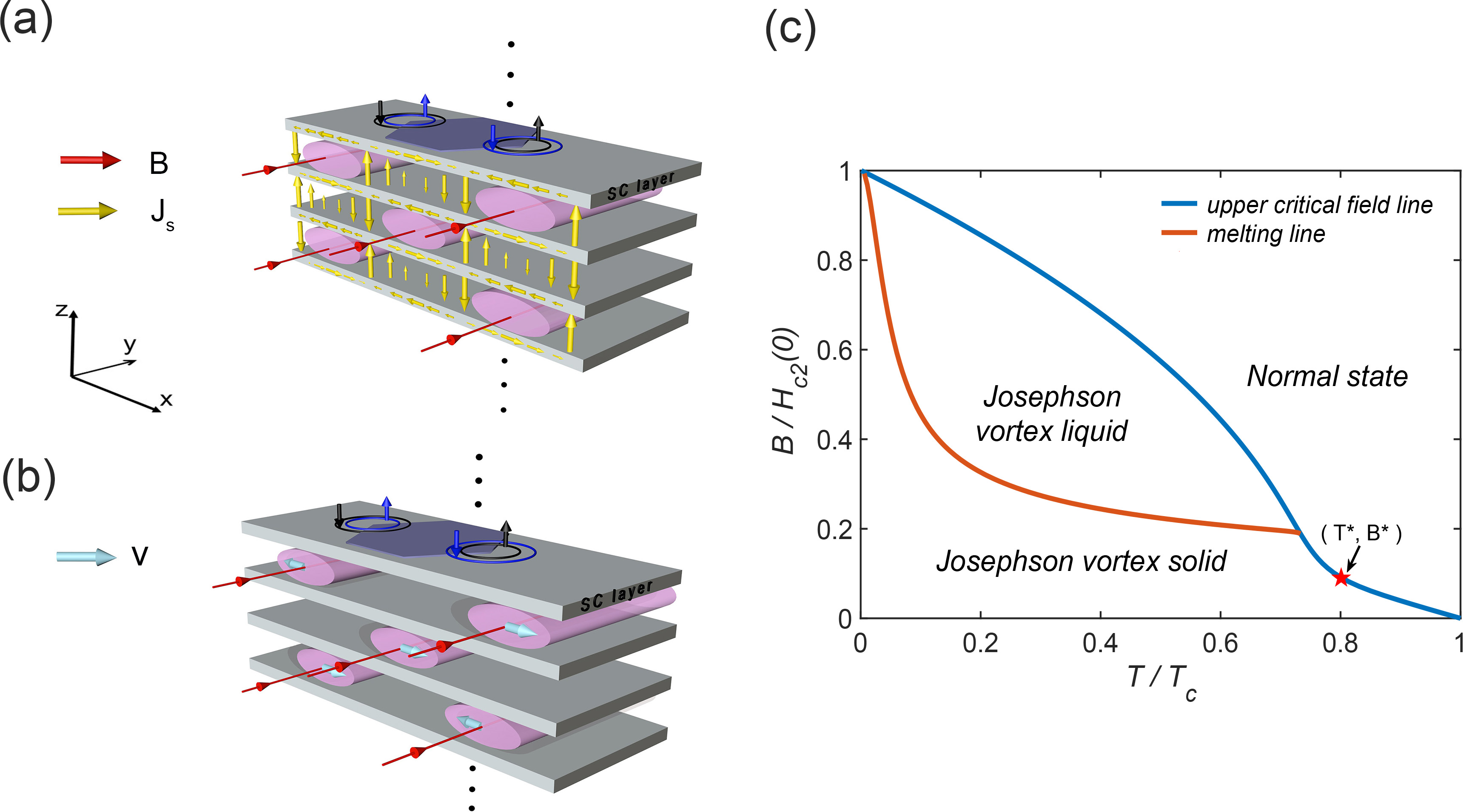}
    \caption{(a) Dense Josephson vortex lattice in layered superconductors. The external magnetic field is oriented in the $y$ direction. The gray parallel boards denote superconducting layers, while the pink elliptical cylinders symbolize the Josephson vortices. The red arrowed lines represent the magnetic flux lines, and the yellow arrows indicate the superconducting current. (b) Josephson vortex liquid. Vortices move randomly in the in-plane directions with the velocity represented by light blue arrows. (c) Phase diagram for the layered superconductors. The blue and orange lines denote the superconducting transition line and the melting line of the Josephson vortex solid, respectively. The states above and below the melting line are schematically depicted by diagrams (a) and (b). The read pentagon represents the crossover point where an upturn of the upper critical field occurs. }
    \label{figure 1}
\end{figure*}

\section{Introduction}\label{sec1}

The interplay between the superconducting phase and magnetic flux has led to significant quantum effects, such as flux quantization and the Fraunhofer diffraction pattern of the supercurrent in Josephson junctions\cite{tinkham2004}. Layered superconductors represent a new frontier for exploring the relationships between superconducting and magnetic flux phases\cite{klemm2011}. In these materials, the Josephson vortex lattice arises from the interplay between interlayer Josephson tunneling and the magnetic vector potential, creating a periodic arrangement of vortices in response to an applied magnetic field \cite{koshelev2013,Clem1990,Bulaevskii1991,Bulaevskii1992,Moler1998,Roditchev2015}. While the melting of the Abrikosov vortex lattice has been extensively studied both theoretically \cite{Nelson1989,Houghton1989,Brandt1989,Brandt1991,Brandt1995,Blatter1993,Blatter1994,Hu1993,onogi1996,vinokur1998,Kierfeld2000} and experimentally \cite{Farrell1991,Kwok19941,Kwok19942,Pstoriza1994,ZeldovMajer1995,Liang1996,Welp1996,Schilling1997,Guillamón2009}, the behavior of the Josephson vortex lattice in layered superconductors under in-plane magnetic fields remains insufficiently explored \cite{Korshunov1991,Korshunov1992,fuhrer1997}, with a notable lack of systematic studies and quantitative criteria.

In conventional superconductors, a magnetic field can induce energy splitting of spin-up and spin-down electrons, leading to Cooper pair destruction when Zeeman energy approaches the superconducting energy gap. Consequently, the upper critical field \(H_{c2}\) is determined by the Pauli limiting field, \(B_{P} = 1.83 T_{c}\), where \(T_{c}\) is the superconducting transition temperature \cite{Chandrasekhar1962,Clogston1962}. The magnetic field can also break Cooper pairs through the orbital effect, characterized by \(B_{orb}\). In the 1960s, Fulde, Ferrell, Larkin, and Ovchinnikov proposed a finite momentum configuration of Cooper pairs, known as the FFLO (Fulde-Ferrell-Larkin-Ovchinnikov) state \cite{Fulde1964,Larkin1965}. This state can withstand stronger magnetic fields, exceeding the Pauli limit. However, successfully observing the FFLO state requires suppressing the influence of the orbital effect, characterized by a large Maki parameter defined as \(\alpha = \sqrt{2} B_{orb}/B_{P}\). To date, evidence of the FFLO state has only been found in heavy fermion systems \cite{Matsuda2007,Bianchi2003,Kakuyanagi2005,Kumagai2006} and organic superconductors \cite{imajo2022,Wosnitza2018,Singleton_2000,Uji2006}.

A new type of finite-momentum superconducting state was predicted by Liu in bilayer transition metal dichalcogenides (TMDs) under an in-plane magnetic field \cite{Liu2017}. The absence of local spatial inversion symmetry results in strong Ising spin-orbital coupling (SOC), which locks paired spins in out-of-plane directions, suppressing the Zeeman effect of the in-plane magnetic field and resulting in Ising superconductivity with a large in-plane critical magnetic field \cite{Lu2015,Saito2016,Xi2016,Zhang2021}. Consequently, these superconductors exhibit a high in-plane upper critical field, distinguishing them from layered organic superconductors. Additionally, the orbital effect is suppressed due to the thinness of the superconducting layers, restricting electron motion. In Liu's proposed orbital-FFLO state in a bilayer Ising superconductor, the momentum of the Cooper pair is non-zero and oppositely directed, reflecting the influence of the vector potential on the phase factor of the superconductor due to the in-plane magnetic field \cite{Liu2017}. Recent experiments suggest the observation of the orbital-FFLO state in multilayer and bulk Ising superconductors \cite{Devarakonda2021,Cho2021,Cho2023,Wan2023,Zhao2023,Yuan2023,yang2024}. Notably, a recent experiment revealed a remarkable first-order phase transition in multilayer 2H-NbSe$_{2}$, characterized by a sudden jump in the superconducting energy gap observed via scanning tunneling spectroscopy, possibly corresponding to a transition to the orbital-FFLO state \cite{cao2024}.

In this paper, we present a detailed analysis of the upper critical boundary and superconducting state of layered Ising superconductors. We introduce a novel configuration of the finite-momentum superconducting state that differs significantly from previous works. Importantly, this configuration becomes increasingly energetically favorable with the addition of layers. We investigate the states before and after the identified first-order phase transition, proposing that the Josephson vortex lattice and its melting play a crucial role in this process. Our work calculates the melting line of the Josephson vortex lattice and provides a qualitative framework for interpreting experimental findings, thereby offering new insights into the behavior of layered Ising superconductors.

The structure of this paper is organized as follows. In Sec. \ref{sec2}, we introduce the phase diagram of the bulk layered superconductors described by our theory. We also present the model that serves as the foundation for our investigation. In Sec. \ref{sec3}, we calculate the upper critical field of the bulk systems with different parameters. In Sec. \ref{sec4}, we study in detail the structure of the dense Josephson vortex lattice presents in the system, and examine the melting process, which is closely associated with the first-order phase transition. Finally, in Sec. \ref{sec5}, we summarize our findings and compare them with existing theories in the literature. We also offer predictions regarding phenomena that can be observed experimentally.  \\

\section{Model}\label{sec2}
We propose that the first-order phase transition observed experimentally corresponds to the melting of the Josephson vortex lattice into the Josephson vortex liquid. Figures \ref{figure 1}(a) and \ref{figure 1}(b) illustrate the distinct states of the Josephson vortex: in the solid state, vortices are arranged in a triangular lattice \cite{Bulaevskii1991}, while in the liquid state, they exhibit random motion primarily in the $x$ direction. The random movement of vortices in the liquid phase results in the magnetic field being effectively averaged over time to uniformly penetrate the superconductor. The blue line in Fig. \ref{figure 1}(c) represents the in-plane upper critical field for bulk layered superconductors, beyond which the order parameter diminishes to zero, signaling a transition to the normal state. This line shows a linear behavior near the superconducting transition temperature \( T_{c} \), but exhibits an upturn as the temperature approaches the crossover point \( (T^{*}, B^{*}) \), marked by a red pentagram in Fig. \ref{figure 1}(c). The orange line indicates the theoretically calculated melting line, notably not connecting to the crossover point \( (T^{*},B^{*}) \) on the upper critical field line.  \\

For layered Ising superconductors, the anisotropy is typically very large due to the weak Josephson coupling between adjacent layers. This significant anisotropy results in a discontinuous order parameter in the out-of-plane direction, distinguishing these layered superconductors from those described by the continuous three-dimensional Ginzburg-Landau model. The strong Ising spin-orbit coupling (SOC) effectively locks the electron spins in the out-of-plane direction, thereby suppressing the Zeeman effect of the in-plane magnetic field \cite{Xi2016}. Consequently, we can disregard the spin degree of freedom of Cooper pairs, as it does not contribute to the free energy of the system. Layered Ising superconductors can be modeled using the well-known Lawrence-Doniach (LD) model, with the free energy expressed as follows \cite{Bulaevskii1992}:

\begin{widetext}
\begin{equation}\label{eq1}
\begin{aligned}
F= & \frac{H_{c}^2(T)}{4\pi} \sum_{l} \int d^2 \mathbf{r} \left \{  \int^{lD+d/2}_{lD-d/2} dz \left [ \xi_{ab}^{2}(T) \left | \left ( \mathbf{\nabla}_{||} -i \frac{2\pi}{\Phi_{0}} \mathbf{A}_{||}(\mathbf{r},z) \right ) \Psi_{l}(\mathbf{r},z) \right | ^2 + \xi_{c}^{2}(T) \left | \left ( \frac{\partial}{\partial z} -i \frac{2\pi}{\Phi_{0}} A_{z}(\mathbf{r},z) \right ) \Psi_{l}(\mathbf{r},z) \right | ^2 \right. \right. \\
& \left. \left. -  \left | \Psi_{l}(\mathbf{r},z) \right | ^2 + \frac{1}{2} \left | \Psi_{l}(\mathbf{r},z) \right | ^4  \right ]  - \frac{\xi_{ab}^{2}(T)}{\lambda_{J}^2} d \left [ \Psi^{*}_{l} \left ( \mathbf{r},lD+\frac{d}{2} \right ) \Psi_{l+1}\left ( \mathbf{r},(l+1)D-\frac{d}{2} \right ) e^{-i \chi_{l,l+1}(\mathbf{r})} +c.c. \right ] \right \} \\
& + \int d^3 \mathbf{r} \frac{B^2(\mathbf{r},z)}{8 \pi}, \\
\chi_{l,l+1}&(\mathbf{r})  = \frac{2\pi}{\Phi_{0}} \int_{lD+d/2}^{(l+1)D-d/2} dz A_{z}(\mathbf{r},z).
\end{aligned}
\end{equation}
\end{widetext}
In Eq. (\ref{eq1}), \( H_c \) represents a characteristic scale of the critical field, expressed as \( H_c = \frac{\Phi_0}{2\sqrt{2}\pi \xi_{ab} \lambda_{ab}} \). We assume that the superconducting layers possess a finite thickness \( d \) and are separated by an interlayer distance \( D \). The parameters \( \xi_{ab} \) and \( \xi_{c} \) denote the in-plane and out-of-plane coherence lengths of a single superconducting layer, respectively. The coherence length is solely a function of temperature and can be expressed as \( \xi = \xi(0)(1 - \frac{T}{T_{c0}})^{-1/2} \), where \( \xi(0) \) corresponds to the coherence length at absolute zero, and \( T_{c0} \) is the transition temperature of an individual layer. It is important to distinguish the coherence length of a single layer from that of the entire system, denoted as \( \xi^{s} \). The temperature dependence of \( \xi^{s} \) is given by \( \xi^{s} = \xi(0)(1 - \frac{T}{T_{c}})^{-1/2} \), where \( T_{c} \) is the superconducting transition temperature of the system. The components of the magnetic vector potential are represented as \( A_{||} \) and \( A_{\perp} \) for the in-plane and out-of-plane directions, respectively. The Josephson length, denoted as \( \lambda_{J} \), is defined by \( \lambda_{J} = \gamma D \), where \( \gamma \) is the anisotropy factor of the system, satisfying the relation \( \gamma = \frac{\xi^{s}_{ab}}{\xi^{s}_{c}} = \frac{\lambda^{s}_{c}}{\lambda^{s}_{ab}} \). The order parameter of the \( l \)-th layer, \( \Psi_{l} \), is measured in units of the magnitude \( \Psi_{0} \), defined as \( \Psi_{0} = -\frac{\alpha}{\beta} \), where \( \alpha \) and \( \beta \) are the coefficients of the second-order and fourth-order terms of the original Ginzburg-Landau model \cite{tinkham2004}. Finally, the term associated with the interlayer Josephson coupling energy includes the Peierls phase \( \xi_{l,l+1} \) between the \( l \)-th and \( (l+1) \)-th layers, which ensures gauge invariance in the presence of a magnetic field.

\begin{figure*}
    \centering
    \includegraphics[width=15cm,height=10cm]{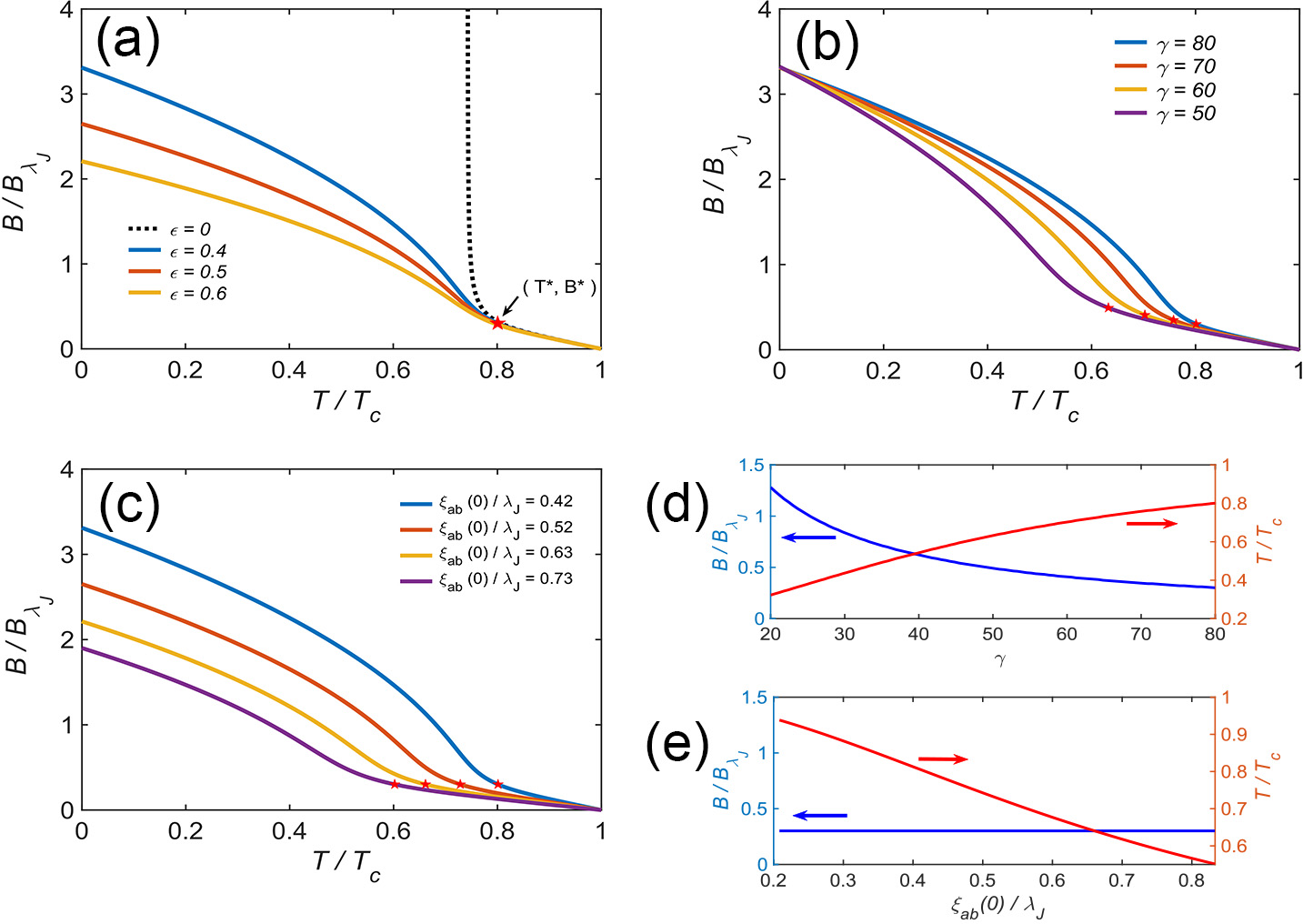}
    \caption{(a) In-plane upper critical fields of bulk layered superconductors with varying ratios \( \epsilon \) of superconducting layer thickness \( d \) to interlayer spacing \( D \). The anisotropy \( \gamma = 80 \) and in-plane coherence length \( \xi_{ab}(0)/\lambda_{J} = 0.42 \) remain constant. The magnetic field scale \( B_{\lambda_{J}} \) is defined as \( B_{\lambda_{J}} = \frac{\Phi_{0}}{\lambda_{J} D} \). (b) In-plane upper critical fields with varying anisotropy \( \gamma \), while keeping the in-plane coherence length \( \xi_{ab}(0)/\lambda_{J} = 0.42 \) and the ratio \( \epsilon = 0.4 \) constant. (c) In-plane upper critical fields with different in-plane coherence lengths, with \( \gamma = 80 \) and \( \epsilon = 0.4 \) held constant. Red pentagons represent crossover points. (d) Magnetic field and temperature of the crossover point as a function of anisotropy \( \gamma \), with \( \xi_{ab}(0)/\lambda_{J} = 0.42 \) and \( \epsilon = 0.4 \) fixed. (e) Magnetic field and temperature of the crossover point as a function of in-plane coherence length \( \xi_{ab}(0)/\lambda_{J} \), maintaining \( \gamma = 80 \) and \( \epsilon = 0.4 \). All figures maintain a constant interlayer spacing of \( D = 1.2\, nm \).}
    \label{figure 2}
\end{figure*}

\section{Upper Critical Field of the bulk layered Ising superconductors}\label{sec3}
Near the upper critical field of the superconductor, the magnetic field is assumed to penetrate uniformly, directed along the \( y \) direction as \( \mathbf{B} = B \mathbf{\hat{y}} \). We choose a gauge such that the magnetic vector potential has only an \( x \) component: \( \mathbf{A} = Bz \mathbf{\hat{x}} \) and \( A_{\perp} = 0 \). Consequently, the Peierls phase \( \chi_{l,l+1} = 0 \). To minimize the free energy, it follows from Eq. (\ref{eq1}) that \( \Psi_{l}(\mathbf{r},z) \) must be independent of the \( y \) and \( z \) coordinates. For bulk materials, we can neglect boundary effects, allowing us to represent the modulus of the order parameter for all layers by the same function \( f \). Thus, the order parameter takes the form \( \Psi_{l}(x) = f(x)e^{i\varphi_{l}(x)} \), where \( \varphi_{l} \) is the phase of the order parameter in the \( l \)-th layer. Substituting the forms of \( \mathbf{A} \) and \( \Psi_{l} \) into Eq. (\ref{eq1}) and omitting the constant magnetic energy term, we simplify Eq. (\ref{eq1}) to:

 \begin{equation}\label{eq2}
\begin{aligned}
F= & \frac{H_{c}^2}{4\pi} \sum_{l} \int d^2 \mathbf{r} \left \{  \int^{lD+d/2}_{lD-d/2} dz \left [ \xi_{ab}^{2}(T) \left (  \left | \frac{d}{d x} f \right |^2  \right. \right. \right. \\
& \left. \left. \left. + \left | \left ( \frac{d \varphi_{l}}{d x}-\frac{2\pi}{\Phi_{0}} Bz \right)  f \right |^2 \right ) - \left |f \right | ^2 + \frac{1}{2} \left | f \right | ^4  \right ] \right. \\
&  \left.  -  2\frac{\xi_{ab}^{2}(T)}{\lambda_{J}^2}d \,  cos(\varphi_{l+1}-\varphi_{l}) \left |f \right | ^2 \right \}. \\
\end{aligned}
\end{equation}
Assume the phase \( \varphi_{l} \) is given by \( \varphi_{l} = Q_{l}x \), where \( Q_l \) represents the mechanical momentum of Cooper pairs in the \( l \)-th layer. From Eq. (\ref{eq2}), if \( Q_{l} \) is constant across all layers (with the uniform superconducting state being a special case when \( Q_{l} = 0 \)), the in-plane kinetic energy term would become very large as the number of layers increases. To minimize the term \( \frac{d \varphi_{l}}{dx} - \frac{2\pi Bz}{\Phi_{0}} \) in Eq. (\ref{eq3}) for lower free energy, we propose that \( Q_l \) should depend on the layer number while remaining independent of the \( z \) coordinate within a single superconducting layer, specifically, \( Q_{l} = \frac{2\pi B z_{l}}{\Phi_{0}} \), where \( z_{l} \) is an undetermined constant. By applying this ansatz into Eq. (\ref{eq2}) and minimizing the free energy with respect to \( z_{l} \), we obtain \( z_{l} = lD \), which corresponds to the \( z \)-axis coordinate at the center of the superconducting layers. This configuration can be viewed as an extension of the finite momentum pairing proposed by Liu in bilayer TMD systems. The free energy then transforms to:

\begin{equation}\label{eq3}
\begin{aligned}
    F= & \frac{H_{c}^2}{4\pi} d \sum_{l} \int d^2 \mathbf{r} \left \{ \xi_{ab}^{2}(T) \left (  \left | \frac{df}{d x}  \right |^2  + \frac{d^2}{12} \left| \frac{2\pi}{\Phi_{0}}B f \right|^2 \right ) \right.   \\
& \left.  - \left |f \right | ^2 + \frac{1}{2} \left | f \right | ^4  - 2\frac{\xi_{ab}^{2}}{\lambda_{J}^2} \, cos\left (\frac{2\pi}{\Phi_{0}}BDx \right ) \left |f \right | ^2 \right \}. \\
\end{aligned}
\end{equation}
First, we examine the special case without a magnetic field. As the temperature approaches the transition temperature \( T_c \) of the system, the spatial derivative \( \frac{df}{dx} \) decreases more rapidly than the modulus \( f \), indicating that \( \frac{df}{dx} \) is a higher-order infinitesimal. By neglecting the \( \frac{df}{dx} \) term and the fourth-order term in Eq. (\ref{eq3}), we can derive the condition for minimum free energy with respect to \( f \). This leads to the relation between the transition temperature of a single layer and that of the entire system: \( T_{c} = \left( \frac{2 \xi_{ab}^2(0)}{\lambda_{J}^2} + 1 \right) T_{c0} \). This result implies that the interlayer Josephson coupling enhaces the transition temperature of the system, which is always greater than that of a single layer. To further simplify this equation, we perform a dimensionless treatment. We define the coordinate scale \( L_{0} = \frac{\Phi_{0}}{BD} \) and introduce the dimensionless length \( \Bar{x} = \frac{x}{L_{0}} \). Eq. (\ref{eq3}) can then be rewritten as:

\begin{equation}\label{eq4}
\begin{aligned}
F= & \frac{H_{c}^2}{4\pi}dL_{x}L_{y} \sum_{l} \int_{0}^{1} d\Bar{x} \left \{ \frac{\xi_{ab}^{2}(T)}{L_{0}^2} \left(  \left | \frac{df}{d \Bar{x}}  \right |^2  + \frac{\pi^2}{3} \epsilon^2  \left |f \right | ^2 \right) \right.   \\
& \left.  - \left |f \right | ^2 + \frac{1}{2} \left | f \right | ^4  - 2\frac{\xi_{ab}^{2}(T)}{\lambda_{J}^2} \, cos\left (2\pi \Bar{x} \right ) \left |f \right | ^2 \right \}. \\
\end{aligned}
\end{equation}
\( L_{x} \) and \( L_{y} \) denote the dimensions of the system in the \( x \) and \( y \) directions, respectively. After performing a coordinate transformation, the modulus \( f \) exhibits a periodicity of 1 in the \( x \) direction. The parameter \( \epsilon \) represents the ratio of the superconducting layer thickness \( d \) to the interlayer spacing \( D \). By varying the free energy with respect to the modulus \( f \) and neglecting the fourth-order term, we derive the eigenvalue equation corresponding to the condition for minimum free energy:

\begin{equation}\label{eq5}
\begin{aligned}
& \hat{L} f = E f,    \\
& \hat{L} =  \frac{d^2}{d \Bar{x}^2} +  2 \frac{L_{0}^{2}}{\lambda_{J}^{2}}  cos \left (2\pi \Bar{x} \right ), \\
& E = \frac{\pi^2}{3}\epsilon^2-  \frac{L_{0}^{2}}{\xi_{ab}^{2}(0)} \left(1- \frac{T}{T_{c0}} \right). \\
\end{aligned}
\end{equation}
This second-order differential equation can be solved numerically. To determine the superconducting phase boundary, we calculate the maximum eigenvalue of the operator \(\hat{L}\), which corresponds to the highest temperature for a given magnetic field. We define a magnetic field scale \(B_{\lambda_{J}} = \frac{\Phi_{0}}{\lambda_{J} D}\). Figure \ref{figure 2}(a) shows that when the ratio of the superconducting layer thickness to the interlayer spacing is zero (\(\epsilon = 0\)), the upper critical field diverges near the crossover point, indicating a transition from three-dimensional to two-dimensional behavior. This non-physical line results from completely neglecting the thickness of the superconducting layers. When \(\epsilon\) is non-zero, this divergence is eliminated \cite{Schneider1993}. At high temperatures, the coherence length in the vertical direction \(\xi^{s}_{c}\) is sufficiently long, making variations in the superconducting layer thickness negligible. As the temperature decreases and \(\xi^{s}_{c}\) approaches the interlayer spacing \(D\), differences between various ratios gradually emerge, causing these lines to separate. Interestingly, at lower temperatures, a larger \(\epsilon\) results in a lower upper critical field. This phenomenon occurs because the in-plane magnetic field causes the electrons to rotate in the \(x-z\) plane, thereby suppressing superconductivity.

In Figs. \ref{figure 2}(b) and \ref{figure 2}(c), we study the effects of anisotropy \(\gamma\) and in-plane coherence length \(\lambda_{ab}(0)\) on the upper critical field. We find that anisotropy \(\gamma\) simultaneously affects both the magnetic field and temperature of the crossover points, marked by red pentagons. As shown in Fig. \ref{figure 2}(d), increasing anisotropy \(\gamma\) shifts the crossover point to higher temperatures and lower magnetic fields. In contrast, the in-plane coherence length only influences the temperature, while the magnetic field at these crossover points remains unchanged, as depicted in Fig. \ref{figure 2}(e). It is important to note that these calculations apply to bulk layered Ising superconductors. When the number of layers is reduced to just a few, the situation changes significantly \cite{Liu2017, Yuan2023}. In such cases, alternative configurations of Cooper pair momenta may arise, potentially leading to more energetically favorable states in certain regions of the phase diagram. In the Supplementary Materials, we provide systematic comparison between superconductors of different layer numbers.

\section{Josephson vortex lattice and its melting}\label{sec4}
Under an in-plane magnetic field, Josephson vortices form in the non-superconducting regions between the superconducting layers. These vortices can exhibit various phases, and their movement leads to energy dissipation and finite resistance, significantly affecting the properties of layered superconductors. In this section, we will discuss the Josephson vortex lattice and its melting in detail.

In type II superconductors, magnetic flux lines penetrate the material and form vortices when the external magnetic field exceeds the lower critical field \(H_{c1}\). In layered superconductors subjected to an in-plane magnetic field, Josephson vortices emerge in the interlayer regions. These vortices possess normal-state cores with dimensions comparable to the coherence lengths \(\xi^{s}_{ab}\) and \(\xi^{s}_{c}\) in the in-plane and out-of-plane directions, respectively. As the magnetic field strengthens, these vortices organize into a triangular lattice, with the lattice periods in the \(x\) and \(z\) directions varying as a function of the magnetic field.

\begin{figure}
    \centering
    \includegraphics[width=8.5cm]{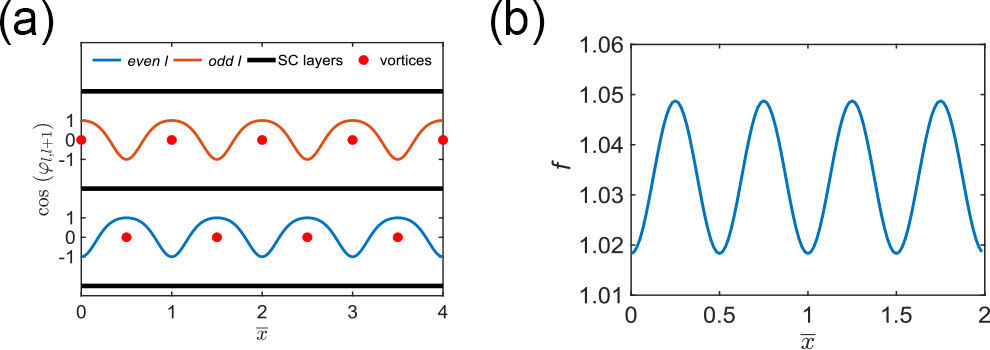}
    \caption{(a) The cosine of the phase difference between neighboring superconducting layers is illustrated. The blue line represents the phase difference between odd and even layers, while the orange line shows the phase difference between even and odd layers. The red circles indicate the cores of the Josephson vortices, which play a crucial role in the vortex dynamics. (b) The spatial variation of the amplitude of the order parameter is shown. Both figures are calculated at a magnetic field of \( B = 0.5  B_{\lambda_{J}} \) and a temperature of \( T = 0.6 T_{c} \), highlighting the relationship between the phase differences and the order parameter amplitude in the context of vortex dynamics.}
    \label{figure 3}
\end{figure}

Our investigation focuses on a densely packed Josephson vortex lattice, which we hypothesize undergoes melting. The magnetic field within the superconductor is non-uniform, expressed as \(B_v = B + \Delta B(x)\), where \(B\) is the average field component and \(\Delta B\) represents a periodically modulated component. However, at high field strengths, the distance between vortices decreases, making \(\Delta B\) negligible compared to \(B\). Thus, we consider only the average magnetic field and express the magnetic vector potential as \(A_{x} = Bz\). From Eq. (\ref{eq3}), we can extract the free energy \(F_{\varphi}\) associated with the phase \(\varphi_{l}\):

\begin{equation}\label{eq6}
\begin{aligned}
F_{\varphi}= & \frac{H_{c}^2 }{4\pi} \xi_{ab}^2 (T) \sum_{l} \int d^2 \mathbf{r} \left \{  \int^{lD+d/2}_{lD-d/2} dz \left | \left ( \frac{d \varphi_{l}}{d x}-\frac{2\pi}{\Phi_{0}} Bz \right)  f \right |^2  \right. \\
&\left.  - \frac{2}{\lambda_{J}^{2}} d \, cos(\varphi_{l+1}-\varphi_{l}) \left | f \right |^2 \right \}. \\
\end{aligned}
\end{equation}
We assume the phase of the order parameter \(\varphi_{l}\) takes the form \(\varphi_{l} = Q_{l}x + P_{l}(x) + C_{l}\). Here, \(P_{l}(x)\) represents a small periodic oscillation term in the \(x\) direction, and \(C_{l}\) is a constant that depends on the layer. We will substitute this expression into Eq. (\ref{eq6}) to proceed with our analysis,

 \begin{equation}\label{eq7}
 \begin{aligned}
 F_{\varphi}=& \frac{H_{c}^2 }{4\pi} \xi_{ab}^2 (T) d  \sum_{l} \int d^2 \mathbf{r} \left \{ \left | f \right |^2 \left [  \left( \frac{d P_{l}}{d x} \right )^2  \right. \right. \\
&  \left. \left. - \frac{2}{\lambda_{J}^{2}} \, cos \left ( \frac{2\pi x}{L_{0}} +P_{l,l+1} +C_{l,l+1} \right ) \right]   \right \}.  \\
 \end{aligned}
 \end{equation}
In the equation above, we define the phase difference between neighboring layers \(l\) and \(l+1\) as \(P_{l,l+1}(x) = P_{l+1}(x) - P_{l}(x)\) and the difference in the constant part as \(C_{l,l+1} = C_{l+1} - C_{l}\). The total phase difference is then given by \(\varphi_{l,l+1}(x) = \frac{2\pi x}{L_{0}} + P_{l,l+1}(x) + C_{l,l+1}\). The term \(\frac{\pi^2 \epsilon^2 |f|^2}{3}\), which arises from the finite thickness, does not affect the phase and is excluded from the phase-related free energy \(F_{\varphi}\). The Josephson current is expressed as \(J = -\frac{\partial F}{\partial A}\), indicating that the current density \(J_{l,l+1}\) between layers \(l\) and \(l+1\) is proportional to \(\cos(\varphi_{l,l+1})\). In a densely packed triangular Josephson vortex lattice, the requirement for periodicity dictates that the current density in different unit cells must be equal, which imposes a constraint on \(\varphi_{l,l+1}\). For layers \(l=2n\) and \(l=2n+1\), periodicity leads to the conditions \(\varphi_{2n-1,2n}(x + L_{0}) = \varphi_{2n-1,2n}(x) + 2\pi\) and \(\varphi_{2n,2n+1}(x + L_{0}/2) = \varphi_{2n-1,2n}(x) + 2\pi\). Consequently, \(P_{l,l+1}\) and \(C_{l,l+1}\) satisfy the following conditions:

\begin{equation}\label{eq8}
\begin{aligned}
& P_{2n-1,2n}(x+L_{0}) = P_{2n-1,2n}(x), \\
& P_{2n,2n+1}(x+L_{0}/2) = P_{2n-1,2n}(x), \\
& C_{2n,2n+1}=C_{2n-1,2n}+\pi.   \\
\end{aligned}
\end{equation}
We consider a specific case where \(C_{2n-1,2n} = 0\) and \(C_{2n,2n+1} = \pi\). Additionally, we define \(C_{2n} = n\pi\) and \(C_{2n+1} = (n+1)\pi\). This choice, while specific, does not lose generality since it captures the essential features of the oscillation and phase relationships in the system. To determine the form of the oscillation term \(P_{l}\), we will vary the free energy \(F_{\varphi}\) in Eq. (\ref{eq7}) with respect to \(P_{2n}\) and \(P_{2n+1}\),

\begin{equation}\label{eq9}
\begin{aligned}
 \frac{\partial^{2} P_{2n-1}}{dx^2}+& \frac{1}{\lambda_{J}^{2}} \left [  sin \left( \frac{2\pi x}{L_{0}} + P_{2n-2,2n-1} +\pi \right) \right. \\
& \left. - sin \left( \frac{2\pi x}{L_{0}} + P_{2n-1,2n} \right) \right ] =0 , \\
 \frac{\partial^{2} P_{2n}}{dx^2}+ & \frac{1}{\lambda_{J}^{2}}   \left [  sin \left( \frac{2\pi x}{L_{0}} + P_{2n-1,2n} \right) \right. \\
& \left. - sin \left( \frac{2\pi x}{L_{0}} + P_{2n,2n+1} +\pi \right) \right ] =0 .  \\
\end{aligned}
\end{equation}
To derive an analytical solution for $P_{l}$ ($l$ includes both odd and even cases), we neglect $P_{l-1,l}$ in sine terms, resulting in a compact equation for $P_{l}$:

\begin{equation}\label{eq11}
P_{l}=(-1)^{l} \frac{L_{0}^2}{2 \pi^2 \lambda_{J}^2} sin \left( \frac{2\pi x}{L_{0}} \right). \\
\end{equation}
Numerical solutions confirm that this simple approximation performs well across most physically relevant parameter regimes, as detailed in the Supplementary Materials. We find that \( P_{2n-2,2n-1} = P_{2n,2n+1} = -P_{2n-1,2n} \) in Eq. (\ref{eq9}). This specific relationship among the \( P_l \) terms relaxes the condition for the validity of Eq. (\ref{eq11}), which no longer requires \( P_{l,l+1} \ll 2\pi x / L_{0} \). Instead, it necessitates that \( \cos\left(\frac{L_{0}^2}{\pi^2 \lambda_{J}^2} \sin\left(\frac{2\pi x}{L_{0}}\right)\right) \sim 1 \). Thus, we propose that the criterion \( \cos\left(\frac{L_{0}^2}{\pi^2 \lambda_{J}^2}\right) > 0.9 \) establishes the minimum magnetic field required for the expression of \( P_{l} \) in Eq. (\ref{eq11}). Ultimately, this leads us to the form of the phase of the order parameter,

\begin{equation}\label{eq13}
\begin{aligned}
&\varphi_{l}(x)=Q_{l}x+S_{l}\pi+(-1)^{l}\frac{L_{0}^2}{2\pi^2 \lambda_{J}^2} sin \left ( \frac{2\pi x }{L_{0}} \right ) \\
&S_{l}=\begin{cases}\frac{l}{2} & \text { if } l \text { is even } \\ \frac{l+1}{2} & \text { if } l \text { is odd }\end{cases}
\end{aligned}
\end{equation}
The cosine values of the phase differences between neighboring superconducting layers are shown in Fig. \ref{figure 3}(a). This result aligns with the findings in \cite{Bulaevskii1991}, indicating a consistent behavior in the phase relationships. By substituting the form of the phase \( \varphi_{l} \) back into Eq. (\ref{eq2}) and transforming \( x \) by setting \( \bar{x} = x/L_{0} \), we can derive the complete expression for the free energy,

\begin{equation}\label{eq14}
\begin{aligned}
F= & \frac{H_{c}^2}{4\pi} dL_{x}L_{y} \sum_{l} \int_{0}^{1} d\Bar{x} \left \{  \frac{\xi_{ab}^2}{L_{0}^2} \left( \left| \frac{\partial f}{\partial \Bar{x}} \right |^2 + \frac{\pi^2}{3}  \epsilon^2  \left |f \right | ^2 \right)  \right.  \\
&  \left.  +\left( \frac{\xi_{ab} L_{0}}{\pi \lambda_{J}^2} \right) ^2  \left[ \frac{3}{2} cos (4 \pi \Bar{x}) - \frac{1}{2} \right] \left |f \right | ^2 -\left |f \right | ^2 + \frac{1}{2} \left |f \right | ^4  \right \} \\
\end{aligned}
\end{equation}
Eq. (\ref{eq14}) represents the free energy of the densely packed Josephson vortex lattice. By minimizing this free energy, we can determine the spatial variation of the modulus of the order parameter, as illustrated in Fig. \ref{figure 3}(b). The modulus is calculated at a magnetic field of \( B = 0.5 B_{\lambda_{J}} \) and a temperature of \( T = 0.6 T_{c} \). Notably, the periodicity of the modulus is half that of the Josephson vortices, reflecting the interactions within the lattice.

Our heuristic physical interpretation of the first-order transition in the superconducting phase is as follows: below the transition line, magnetic vortices are arranged in a lattice structure. At the transition line, the Josephson vortex lattice melts, resulting in the random motion of these vortices, which can be characterized as a Josephson vortex liquid. While the scale of this random motion is microscopic, the magnetic field exhibits more uniform characteristics at the macroscopic level, aligning with the scenario described in Sec. (\ref{sec3}).

A straightforward approach might involve comparing the free energy of the finite-momentum state and the dense Josephson vortex lattice, as shown in Eqs. (\ref{eq4}) and (\ref{eq14}), to identify the melting line. However, this is complicated by the latent heat associated with the phase transition\cite{Schilling1997}. During melting, the Josephson vortex lattice absorbs heat from the environment, leading to an inequality between the free energies of these two states. Therefore, we cannot determine the melting line by simply equating these free energies. Instead, we employ the traditional Lindemann criterion to establish the melting line, which provides a reliable framework for analyzing phase transitions in this context.

At finite temperatures, thermal fluctuations induce small vibrations in the vortices, causing them to deviate from their equilibrium positions. This deviation is closely related to the elastic modulus of the Josephson vortex lattice, which reflects the lattice's resistance to deformation under stress. Therefore, we will focus our subsequent analysis on deriving the elastic modulus, as it plays a crucial role in understanding the dynamics of the vortex lattice under thermal influences.

In layered superconductors, small deformations in the $x$ direction, denoted as $U_{l}^{x}(\mathbf{r})$, can lead to changes in the phase of the order parameter. In contrast, deformations in the $y$ direction do not affect the phase distribution when considering only the first-order effects. We also neglect deformations along the $z$ direction for two reasons: first, the presence of superconducting layers strongly restricts such deformations, and second, the phase $\varphi$ remains unchanged under small out-of-plane deformations.

Consequently, an additional term $V_{l}^{x}(\mathbf{r})$ is introduced into the phase of the order parameter, where $V_{l}^{x}(\mathbf{r}) \ll 1$ and $\mathbf{r} = (x, y)$. It is important to note that both $U_{l}^{x}(\mathbf{r})$ and $V_{l}^{x}(\mathbf{r})$ depend on the $x$ and $y$ coordinates, as the deformations in the $x$ direction can vary non-uniformly along the $y$ axis, disrupting the uniformity of the phase. Therefore, the phase of the order parameter $\varphi$ becomes a function of $\mathbf{r}$. For simplicity, we omit the $x$ indices in $V_{l}$ and $U_{l}$. Given the weak deformation, we ensure that other terms remain unaffected, except for the oscillation term $P_{l}$ in Eq. (\ref{eq13}),

\begin{equation}\label{eq15}
\varphi_{l}(\mathbf{r})=Q_{l}x+S_{l}\pi+\Tilde{P}_{l}(x) +V_{l}(\mathbf{r}), \\
\end{equation}
$\Tilde{P}_{l}(x)$ represents the deformed oscillation terms, while $P_{l}$ in Eq. (\ref{eq11}) serves as the zero-order approximation of $\Tilde{P}_{l}(x)$. We assume that $\Tilde{P}_{l}(x)$ retains some characteristics of $P_{l}(x)$ and incorporates its properties. Additionally, $\Tilde{P}_{l}(x)$ satisfies the condition $\Tilde{P}_{l}(x) \ll 1$ for a dense Josephson vortex lattice, similar to $P_{l}(x)$. Substituting the phase expression into Eq. (\ref{eq7}) yields:

\begin{equation}\label{eq16}
 \begin{aligned}
 F_{\varphi}=& \frac{H_{c}^2}{4\pi} \xi_{ab}^2 d  \sum_{l} \int d^2 \mathbf{r}  \left | f \right |^2 \left \{ \left [ \left( \frac{d\Tilde{P}_{l}}{dx} \right)^2 + \left( \frac{dV_{l}}{d\mathbf{r}} \right)^2 \right] \right. \\
 & \left. - \frac{2}{\lambda_{J}^{2}} cos \left [ \frac{2\pi x}{L_{0}} + \pi S_{l,l+1}+\Tilde{P}_{l+1}-\Tilde{P}_{l} +V_{l+1}-V_{l} \right] \right \},  \\
 \end{aligned}
 \end{equation}
The product term of \( \frac{d\Tilde{P}}{dx} \) and \( \frac{dV}{\mathbf{r}} \) vanishes because the integral of the linear term \( \frac{dP}{dx} \) is zero, while the square term remains finite. This can be verified by substituting Eq. (\ref{eq11}) into Eq. (\ref{eq7}). By expanding the cosine term, we obtain:

\begin{equation}\label{eq17}
\begin{aligned}
F_{\varphi}=& \frac{H_{c}^2}{4\pi} \xi_{ab}^2 d  \sum_{l} \int d^2 \mathbf{r} \left | f \right |^2 \left \{ \left [ \left( \frac{d\Tilde{P}_{l}}{dx} \right)^2 + \left( \frac{dV_{l}}{d\mathbf{r}} \right)^2  \right] \right. \\
& \left. + \frac{2}{\lambda_{J}^{2}} sin \left(  \frac{2\pi x}{L_{0}} +\pi S_{l,l+1}+ V_{l+1}-V_{l} \right) \left(\Tilde{P}_{l+1}-\Tilde{P}_{l} \right) \right \}.
\end{aligned}
\end{equation}
The product of the cosine terms in the above equation is eliminated. This is verified by summing the terms for \( l=2n-1 \) and \( l=2n \), given the condition \( V_{l}^{x}(\mathbf{r}), \Tilde{P}_{l} \ll 1 \). By varying \( F_{\varphi} \) with respect to \( \Tilde{P}_{l} \), we obtain the expression for \( \Tilde{P}_{l} \),

\begin{equation}\label{eq18}
\begin{aligned}
\Tilde{P}_{l}=&\frac{(-1)^{l+1}}{(2\pi)^2} \frac{L_{0}^2}{\lambda_{J}^2} \left [ sin \left( \frac{2\pi x}{L_{0}}+V_{l+1}-V_{l} \right) \right. \\
& \left. +sin \left( \frac{2\pi x}{L_{0}}+V_{l}-V_{l-1} \right) \right ],
\end{aligned}
\end{equation}
By substituting the expression back into Eq. (\ref{eq17}), we perform the calculations to obtain the energy of the deformed Josephson vortex lattice:

\begin{figure*}
    \centering
    \includegraphics[width=17cm,height=8cm]{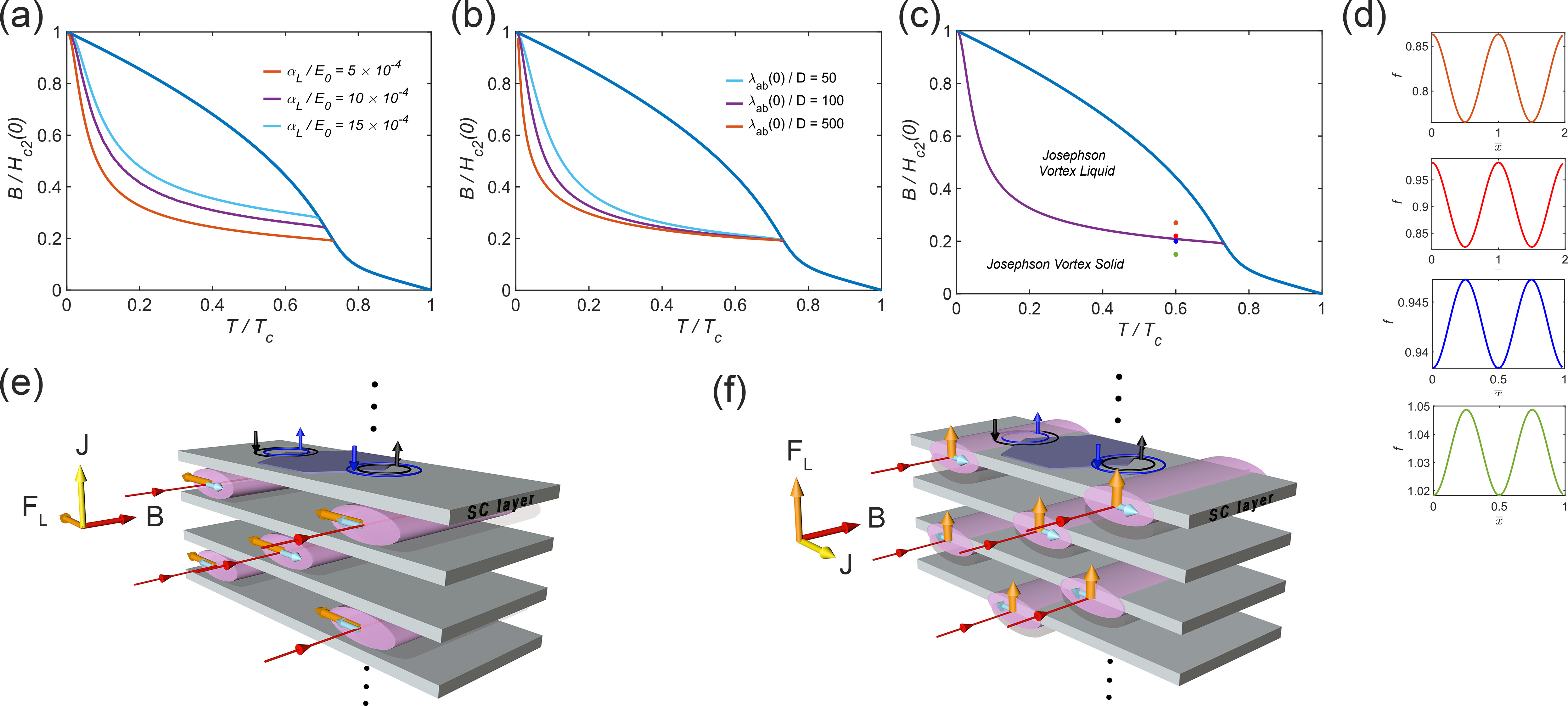}
    \caption{Dependence of the Josephson vortex melting lines and characteristics of Josephson vortex motion under current. (a) Josephson vortex melting line as a function of pinning strength. The orange, purple, and light blue lines correspond to Labusch parameters \(\alpha_{L}/E_{0} = 5 \times 10^{-4}\), \(10 \times 10^{-4}\), and \(15 \times 10^{-4}\), respectively. The in-plane penetration depth is set at \(\lambda_{ab}(0)/D = 100\). (b) Josephson vortex melting line as a function of in-plane penetration depth, with the Labusch parameter fixed at \(\alpha_{L}/E_{0} = 5 \times 10^{-4}\). The orange, purple, and light blue lines represent penetration depths \(\lambda_{ab}(0)/D = 50\), \(100\), and \(500\), respectively. Parameters used: \(T_{c} = 2.7\, K\), \(\gamma = 80\), \(\xi_{ab}(0)/\lambda_{J} = 0.42\) and \(L_{z}/D=10^4\). Geometric parameters \(\kappa\) and the ratio \(R_{\Box}/R_Q\) are fixed at \(\kappa = 0.1\) and \(R_{\Box}/R_Q = 1\), respectively. Lindemann criterion is set to \(c_{L}=0.3\). (c) Phase diagram of the system. Green and blue points are below the melting line, while red and orange points are above it, corresponding to \(B/H_{c2}(0) = 0.50\), \(0.67\), \(0.73\), and \(0.90\). (d) The column of images represent spatial variation of the order parameter modulus at different magnetic field strengths in (c), with \(T/T_{c} = 0.6\). (e) Schematic diagram illustrating the movement of the Josephson vortices under interlayer current. (f) Schematic diagram illustrating the movement of the Josephson vortices under intralayer current. Yellow and orange arrows represent the applied current and the corresponding Lorentz force on the Josephson vortices, while light blue arrows indicate the random motion of these vortices in the in-plane direction.}
    \label{figure 4}
\end{figure*}

\begin{equation}\label{eq19}
\begin{aligned}
F_{\varphi}=& \frac{H_{c}^2}{4\pi} \xi_{ab}^2 d \sum_{l} \int d^2 \mathbf{r} \left | f \right |^2 \left \{ \left ( \frac{dV_{l}}{d\mathbf{r}} \right)^2  \right. \\
& \left. -\left( \frac{L_{0}}{2 \pi \lambda_{J}^{2}} \right)^2 \left [ cos \left( V_{l+1}+V_{l-1}-2V_{l} \right ) +1 \right] \right \}. \\
\end{aligned}
\end{equation}
Since \( V_{l} \) represents a small perturbation in the phase of the order parameter, we can expand \( \cos(V_{l+1}^{x} + V_{l-1}^{x} - 2V_{l}^{x}) \) to second order in \( V_{l} \) to simplify our analysis. This expansion allows us to approximate the behavior of the system under small deformations, facilitating further calculations related to the melting transition of Josephson vortex lattice,

\begin{equation}\label{eq20}
\begin{aligned}
F_{\varphi}=& \frac{H_{c}^2}{4\pi} \xi_{ab}^2 d \sum_{l} \int d^2 \mathbf{r} \left | f \right |^2 \left \{  \left ( \frac{dV_{l}}{d\mathbf{r}} \right)^2  \right. \\
& \left. -\frac{1}{2} \left( \frac{L_{0}}{2 \pi \lambda_{J}^{2}} \right)^2 \left( V_{l+1}+V_{l-1}-2V_{l} \right)^2  \right \}. \\
\end{aligned}
\end{equation}
We then perform a Fourier transformation on the above equation, $V_{l}(\mathbf{r})=1/\sqrt{V} \sum_{k} V_{k} e^{ik_{||} \cdot \mathbf{r}} e^{ik_{z} lD}$, where $V$ is the total volume of the system, and obtain

\begin{equation}\label{eq21}
\begin{aligned}
F_{\varphi}=& \frac{H_{c}^2}{4\pi} \xi_{ab}^2 \epsilon \left | \Bar{f} \right |^2 \sum_{k} \left |V_{k} \right|^{2} \\
& \times \left \{ k_{x}^2 +2 \left( \frac{L_{0}}{2 \pi \lambda_{J}^{2}} \right)^2 \left [ 1-cos\left( k_{z}D \right) \right]^2   \right \}.
\end{aligned}
\end{equation}
In the above equation, we perform the summation over \( k \) within the first Brillouin zone, assuming that in strong magnetic fields, the spatial modulation of the superconducting order parameter amplitude is negligible. For simplicity, we replace the amplitude with an average value derived from Eq. (\ref{eq14}). The small deformations \( U_{l}(\mathbf{r}) \) are related to the phase deformation via \( U_{l}^{x} = (V_{l+1}^{x} - V_{l}^{x})L_{0}/2\pi \), leading to the expression \( U_{k} = V_{k}(e^{ik_{z}D} - 1)L_{0}/2\pi \). By substituting \( V_{k} \) with \( U_{k} \) in Eq. (\ref{eq21}), we obtain:

\begin{equation}\label{eq22}
\begin{aligned}
& F_{\varphi}=\frac{1}{2} \sum_{k} \left[c_{11} k_{x}^{2}+c_{66} k_{y}^{2}+c_{44}\Tilde{k}_{z}^{2} \right] \left |U_{k} \right |^2, \\
& c_{11}=c_{66}=\frac{B^{2} \epsilon}{4\pi \lambda_{ab}^{2} \Tilde{k}_{z}^{2}}\left | \Bar{f} \right |^2, \quad c_{44}=\frac{\Phi_{0}^2 \epsilon}{32 \pi^3 \gamma^{4} \lambda_{ab}^{2}  D^{2}}\left | \Bar{f} \right |^2 .\\
\end{aligned}
\end{equation}
In the above equation, we define \( \Tilde{k}_{z} = 2 \sin(k_{z}D/2)/D \). Here, \( c_{11} \) represents the uniaxial compression modulus, \( c_{44} \) denotes the out-of-plane tilting modulus, and \( c_{66} \) is the in-plane shear modulus \cite{Brandt1977}, all of which arise from the Josephson coupling between adjacent layers. In the case of a dense Josephson vortex lattice, the moduli \( c_{11} \) and \( c_{66} \) are equivalent \cite{koshelev2013}. We utilize the traditional definition of elastic moduli in Eq. (\ref{eq22}).

Once we have determined the elastic moduli of the dense Josephson vortex lattice, we can calculate the vortex melting line using the Lindemann criterion. This criterion states that \( c_{L}^{2} = \left \langle u^2 \right \rangle / L_{0}^{2} \). When the ratio \( c_L \) surpasses a threshold (typically between 0.1 and 0.3) \cite{Nelson1989,Houghton1989,Blatter1993,Blatter1994}, a transition from a vortex solid to a vortex liquid occurs. Although understanding the microscopic melting process poses challenges, the Lindemann criterion serves as an effective empirical tool for qualitatively assessing melting phenomena. To calculate the mean square displacement, we apply the fluctuation-dissipation theorem \cite{Kubo1966}, which connects the system's response to external disturbances with its fluctuations. We define the time-domain correlation function \( K(\tau) \) as follows:

\begin{equation}\label{eq23}
K(\tau)= \frac{1}{2}\left \langle u(t+\tau)u(t) + u(t)u(t+\tau) \right \rangle .
\end{equation}
The displacement of the vortices at time \( t \) is represented by \( u(t) \), and \( \tau \) is the time difference. The angle brackets denote an average over all vortices, reflecting the overall behavior of the vortex lattice. The mean square displacement is calculated from the correlation function by setting \( \tau = 0 \): \( \left \langle u^2 \right \rangle = K(0) = \left \langle u(t) u(t) \right \rangle. \) Choosing \( \tau = 0 \) gives us the fluctuations of the vortices at the same moment, indicating their stability around equilibrium. As temperature increases, thermal fluctuations cause larger deviations from equilibrium, leading to an increase in \( \left \langle u^2 \right \rangle \). Additionally, the Fourier component of the correlation function, \( S(\omega) \), can be derived through Fourier transformation, which helps analyze the system's response to fluctuations.

\begin{equation}\label{eq24}
 S(\omega)=\frac{1}{2\pi} \int_{-\infty}^{\infty} K(\tau) e^{i\omega \tau} d\tau. \\
\end{equation}
From the fluctuation-dissipation theorem, we have

\begin{equation}\label{eq25}
S(\omega) = \frac{\hbar}{2\pi} \mathbf{Im} G(\omega) \, coth \left (\frac{\hbar \omega}{2k_{B}T} \right).
\end{equation}
In the above formula, \( G(\omega) \) represents the response function. We account for both pinning and damping effects between the superconducting layers, which inhibit the motion of the vortices \cite{Schmucker1977}. For simplicity, we assume that the magnitudes of these effects are proportional to the velocity \( \dot{u} \) and the displacement \( u \), respectively. Consequently, the total force \( f \) acting on the vortices can be expressed as follows:

\begin{equation}\label{eq26}
\begin{aligned}
f_{\mathbf{k}} &= \Phi_{\mathbf{k}} u_{\mathbf{k}}  + \alpha_{L} u_{\mathbf{k}} +\eta \dot{u}_{\mathbf{k}} \\
&=  \left [\Phi_{\mathbf{k}} + \alpha_{L} -i\omega \eta \right] u_{\mathbf{k}}.
\end{aligned}
\end{equation}
Here, \(\Phi_{\mathbf{k}}\) is the dynamic matrix defined as \(\Phi_{\mathbf{k}} = c_{11}k_{x}^{2} + c_{66}k_{y}^{2} + c_{44}k_{z}^{2}\). The Labusch parameter \(\alpha_{L}\) represents the restoring force exerted on the vortices per unit displacement, while \(\eta\) is the damping coefficient. We define \(\Phi^{\prime}_{\mathbf{k}} = \Phi_{\mathbf{k}} + \alpha_{L} - i\omega \eta\). The inverse of this effective dynamic matrix, \((\Phi^{\prime}_{\mathbf{k}})^{-1}\), corresponds to the response function \(G(\omega)\) discussed earlier. By combining Eqs. (\ref{eq24}) and (\ref{eq25}), we obtain the mean square displacement \(\left \langle u^2 \right \rangle = \sum_{\mathbf{k}} \left \langle u_{\mathbf{k}}^2(t) \right \rangle\) by setting \(\tau=0\).

\begin{widetext}
\begin{equation}\label{eq27}
\left \langle u^2 \right \rangle =   \frac{\hbar}{2\pi} \mathbf{Im} \left [ \int_{-\infty}^{\infty} d\omega \, coth \left( \frac{\hbar \omega}{2 k_{B}T} \right)  \int_{BZ} \frac{d^3 \mathbf{k}}{(2\pi)^3} \frac{1}{c_{11}k_{x}^{2} + c_{66}k_{y}^{2} + c_{44}\Tilde{k}_{z}^{2} + \alpha_{L} -i\omega \eta} \right],
\end{equation}
\end{widetext}
The integral is performed over the wave vector \(\mathbf{k}\) in the first Brillouin zone. The unit cell of the dense Josephson vortex lattice is a parallelogram in the \(x-z\) plane, defined by the lattice vectors \(a_{1} = (0, L_{0}, 0)\) and \(a_{2} = (0, L_{0}/2, d)\). In the \(y\) direction, the unit cell scale corresponds to the overall system size. Consequently, the reciprocal lattice vectors are given by \(b_{1} = 2\pi (0, 1/L_{0}, -1/2d)\) and \(b_{2} = 2\pi (0, 0, 1/d)\). The wave vector \(k_{y}\) ranges from \([- \pi/L_{y}, \pi/L_{y}]\). If the system size in the \(y\) direction, \(L_{y}\), is much larger than both \(L_{0}\) and \(d\), then \(k_{y}\) will be sufficiently small, allowing us to neglect this term in the denominator of Eq. (\ref{eq27}). The integral with respect to \(k_{y}\) simplifies to \(\int dk_{y}/2\pi = 1/L_{y}\). For \(\omega < 0\), we make the transformation \(\omega \to -\omega\), and Eq. (\ref{eq27}) transforms to:

\begin{equation}\label{eq28}
\begin{aligned}
\left \langle u^2 \right \rangle = &  \frac{\hbar}{2\pi L_{y} \eta}   \int_{0}^{\infty} d\omega \, coth \left( \frac{\hbar \omega}{2 k_{B}T} \right)  \\
& \times \int_{BZ} \frac{dk_{x} k_{z}}{(2\pi)^2} \frac{2 \omega \eta^2}{(c_{11}k_{x}^{2} + c_{44}\Tilde{k}_{z}^{2} + \alpha_{L})^{2}+ \omega^{2} \eta^{2}} \\
= & \frac{\kappa}{\pi^2} \frac{R_{\Box}}{R_Q}  \int_{0}^{\infty} d\omega \, coth \left( \frac{\hbar \omega}{2 k_{B}T} \right) \\
& \times \frac{1}{N_{k}} \sum_{\mathit{\Delta} S_{k}} \frac{2 \omega \eta^2}{(c_{11}k_{x}^{2} + c_{44}\Tilde{k}_{z}^{2} + \alpha_{L})^{2}+ \omega^{2} \eta^{2}}
\end{aligned}
\end{equation}
In the above equation, \(\eta\) represents the viscous drag coefficient in the non-superconducting region between neighboring superconducting layers, expressed as \(\eta = BH_{c2}/c^2 \rho_n\), where \(c\) is the speed of light and \(\rho_n\) is the normal-state resistivity. The geometric parameter \(\kappa\) denotes the ratio of dimensions in the \(y\) and \(z\) directions, defined as \(\kappa = L_{y}/L_{z}\). The sheet resistance \(R_{\Box}\) is given by \(R_{\Box} = \rho_n / L_{z}\), while the quantum resistance \(R_Q\) is defined by fundamental constants \(\hbar/e^2\). Next, \(N_{k}\) and \(S_{k}\) represent the number of divided cells and the area of each cell in the first Brillouin zone, respectively. Equation (\ref{eq28}) should be truncated at a maximum frequency \(\omega_{max}\), given by \(\omega_{max} = c^2 \rho/2\pi^2 \lambda_{ab}^2\) according to \cite{Blatter1993}. To facilitate our analysis, we introduce a stiffness scale \(E_{0}\) defined as \(E_{0} = B_{\lambda_{J}}^2 /2\pi^2 \lambda_{0}^{2}\), where \(\lambda_{0}/D =100 \) is used in our calculations.

In Fig. \ref{figure 4}(a), we investigate the dependence of the Josephson vortex melting line on pinning strength, quantified by the Labusch parameter \(\alpha_L\). The melting line starts above the crossover point on the upper critical field line and initially remains gentle before steepening as the temperature approaches absolute zero. Stronger pinning effectively constrains the vibrations of Josephson vortices, leading to a melting line characterized by stronger magnetic fields and higher temperatures. Fig. \ref{figure 4}(b) analyzes the effect of the in-plane penetration depth \(\lambda_{ab}(0)\) on the melting line, revealing that a smaller penetration depth increases the elastic modulus, reducing the amplitude of the vortices and shifting the melting line to higher temperatures and stronger magnetic fields.

Upon melting, vortices can move randomly along the \(x\) direction, exhibiting uniform characteristics on a macroscopic scale, allowing us to treat the magnetic field as homogeneous. The spatial variation in the modulus of the order parameter can be determined by minimizing the free energy as described in Eq. (\ref{eq4}). Fig. \ref{figure 4}(c) demonstrates that at a fixed temperature of \(T/T_c = 0.6\), an increasing magnetic field crossing the melting line causes a sudden increase in the amplitude of modulus \(f\) oscillations and a doubling of its period. These changes serve as indirect experimental evidence of vortex melting, providing valuable insights into this complex phenomenon.

The proposed experiment predicts distinct transport behaviors of Josephson vortices under varying magnetic fields, depending on whether an intralayer or interlayer current is applied. In Fig. \ref{figure 4}(d), with the current directed interlayer, the Lorentz force acts in the \(-x\) direction, described by the equation \(F_{L} = J \times \Phi_{0}\). This steady external force causes vortices to move directionally along this path, leading to energy dissipation due to viscosity and resulting in non-zero resistance. Furthermore, this resistance increases with the magnetic field, as a stronger field elevates the vortex density.

Conversely, in Fig. \ref{figure 4}(e), when the current is directed along the \(x\) axis, the Lorentz force acts in the \(z\) direction. Here, Josephson vortices exhibit directional motion along the \(z\) direction while also moving randomly in the \(x\) direction, represented by orange and light blue arrows, respectively. However, the interlayer motion is hindered by the superconducting layers, which act as periodically distributed potential barriers. As a result, the rate at which intralayer resistance increases with respect to the magnetic field is significantly slower than that of interlayer resistance. It is anticipated that experimental results will show interlayer resistance reaching its normal value while intralayer resistance remains below its normal value.

\section{Discussion}\label{sec5}
In this article, we explore the superconducting phase boundary and first-order phase transition in bulk layered Ising superconductors, introducing a novel configuration for Cooper pairs' momenta across layers. Notably, our proposed finite-momentum state differs from the conventional FFLO state, where the non-zero momentum arises from Fermi surface mismatches due to the Zeeman effect of the magnetic field. This momentum is physical and gauge-independent, inducing spatially periodic modulation. In contrast, our finite-momentum state is influenced by the magnetic vector potential, which is gauge-dependent. Our analysis relies on a specific gauge where the vector potential has only in-plane components; other gauges would lead to different forms of the order parameter, thus making the finite momentum a non-physical quantity. However, similar to the Aharonov-Bohm effect, the crucial aspect is the phase difference between adjacent superconducting layers, which is gauge-invariant and observable, manifesting as periodic oscillations in the order parameter's modulus.

Furthermore, we do not address the complexities of few-layer Ising superconductors, as their superconducting state significantly differs from that of bulk systems. The upper critical phase boundary in few-layer systems shows two-dimensional Ginzburg-Landau behavior near the transition temperature when the system's thickness is less than the out-of-plane coherence length, and an upturn in the phase boundary can also occur. The superconducting phase of few-layer systems hosts various candidate states; while a uniform superconducting state is favored in weak magnetic fields, the lowest energy state may shift rapidly as the magnetic field strengthens. We propose that at high magnetic fields, the momentum configuration in few layers will converge to the form discussed in Section \ref{sec3}, with the region of this configuration expanding as the number of layers increases. Although Josephson vortices can appear in few-layer superconductors, analyzing the Josephson vortex lattice is challenging due to significant boundary effects in the $z$ direction and the lack of periodicity. Additionally, understanding the mechanism of first-order transitions in few-layer superconductors is complex, potentially involving transitions from a uniform superconducting state to a finite-momentum state or from a Josephson vortex solid state to a vortex liquid state.

The present analysis primarily applies to layered superconductors with weak Josephson coupling. In systems with strong Josephson coupling, the characteristic upturn in the upper critical field may not be observable within the experimentally accessible temperature range. Furthermore, in the case of very strong coupling, conventional Josephson vortices may cease to exist, and the system may instead exhibit vortex structures similar to Abrikosov vortices found in bulk type-II superconductors. This regime of strong coupling and the resulting vortex behavior fall outside the scope of our current study.

Our calculated melting line closely resembles the first-order phase transition line reported in a recent experimental study by Cao et al.\cite{cao2024}, with both lines originating above the crossover point. Moreover, we are aware of the observations of orbital-FFLO and melting transition in a bulk superconducting superlattice\cite{lin2024}. The Josephson vortex melting transition scenario connects our theoretical predictions with experimental observations, offering a unified perspective on the behavior of layered superconductors under in-plane magnetic field.

In conclusion, our study proposes a more plausible finite momentum configuration for bulk layered Ising superconductors under high magnetic fields. We demonstrate that the concept of finite momentum pairing in layered systems is essentially gauge-dependent. Additionally, our work provides new insights into the nature of the first-order phase transition in bulk systems, identifying it as a Josephson vortex melting process. We predict that crossing the first-order phase transition line will lead to an increased oscillation amplitude of the superconducting gap, accompanied by a doubling of the oscillation period. These spatial variations in the gap could potentially be detected using high-resolution STM scanning probes. For melting occurring at magnetic fields around ten teslas, we anticipate a typical oscillation period of tens of nanometers. Notably, the physical picture of vortex melting presented here parallels that observed in high-temperature superconductors, potentially bridging these two domains of superconductivity research.

\section*{Acknowledgements}
Hongyi Yan and Haiwen Liu are grateful to discussions with Chaoxing Liu, Z. Cao and J. F. Lin. This work was financially supported by the National Natural Science Foundation of China (Grants No. 12374037, No. 12174442 and No. 12361141820), the National Key Research and Development Program of China (Grants No. 2022YFA1403100, No. 2022YFA1403103, and No. 2023YFA1406500), and the Strategic Priority Research Program of the Chinese Academy of Sciences (Grant No. XDB28000000), Innovation Program for Quantum Science and Technology (Grant No. 2021ZD0302400), and the Fundamental Research Funds for the Central Universities.

\bibliographystyle{apsrev4-1}
\bibliography{main}
\clearpage

\onecolumngrid
\begin{center}
    \Large\textbf{Supplementary Materials for ``Orbital-FFLO state and Josephson vortex lattice melting in the layered Ising superconductors"}
\end{center}
 \setcounter{page}{1}

\setcounter{section}{0}
\setcounter{figure}{0}
\renewcommand{\thefigure}{S\arabic{figure}}
\renewcommand{\thesection}{S\arabic{section}}

\section{Finite layer Ising superconductor}

In the bulk materials discussed in the main text, we consider the number of layers to be effectively infinite. This approximation allows us to ignore boundary effects and assume a uniform modulus of the order parameter across all layers. However, recent experiments have often focused on systems with a finite number of layers, ranging from a bilayer to several tens of layers. As the number of layers decreases, the finite-momentum state, which depends on our chosen gauge, may not remain energetically favorable throughout the entire phase diagram. This observation points to the potential for states with alternative momentum configurations of Cooper pairs, which may offer lower energy solutions.

For clarity, we introduce two distinct states based on their momentum characteristics: the previously discussed state from the main text will be referred to as the "A state," which includes the finite-momentum aspect. In contrast, the "orbital FFLO" state, characterized by a consistent momentum \( q \) across all layers, will be termed the "B state." In this section, we set aside considerations of Josephson vortices to focus solely on these two states. We aim to compare their respective regions within the phase diagram and explore how these regions shift with changes in the number of layers.

In the "A state," the lack of periodicity in the out-of-plane direction leads to a layer-dependent modulus of the order parameter. This variation necessitates a revision of the free energy expression to accurately reflect the influence of finite layering. The free energy for each state needs to be recalculated to determine which configuration is more energetically favorable under varying conditions,

\begin{equation}\label{eqS11}
F=  \frac{H_{c}^2}{4\pi}dL_{x}L_{y} \sum_{l=1}^{N} \int_{0}^{1} d\Bar{x} \left \{ \frac{\xi_{ab}^{2}(T)}{L_{0}^2} \left(  \left | \frac{df_{l}}{d \Bar{x}}  \right |^2  + \frac{\pi^2}{3} \epsilon^2  \left |f_{l} \right | ^2 \right)  - \left |f_{l} \right | ^2 + \frac{1}{2} \left | f_{l} \right | ^4  - 2\frac{\xi_{ab}^{2}(T)}{\lambda_{J}^2} \, cos\left (2\pi \Bar{x} \right ) f_{l} f_{l+1} \right \}.
\end{equation}
$N$ denotes the total number of layers in the system. In the B states, the superconducting phase across all layers is uniformly expressed as $\varphi_{l} = qx$, independent of the layer index. This uniformity eliminates spatial modulation in the Josephson coupling term described in Equation (\ref{eq2}), resulting in a consistent modulus of the order parameter within each layer. Assume the order parameter takes the form $\Phi_{l}(x) = f_{l}e^{iqx}$. Consequently, the free energy of the B state is expressed as follows:

\begin{equation}\label{eqS12}
F=  \frac{H_{c}^2}{4\pi}dL_{x}L_{y} \sum_{l=1}^{N} \left \{ \frac{\xi_{ab}^{2}(T)}{L_{0}^{2}} \left [ \left( qL_{0}-2\pi \Bar{l} \right)^{2}  + \frac{\pi^2}{3} \epsilon^{2} \right] \left |f_{l} \right | ^2 - \left |f_{l} \right | ^2 + \frac{1}{2} \left | f_{l} \right | ^4  - 2\frac{\xi_{ab}^{2}(T)}{\lambda_{J}^2} f_{l} f_{l+1} \right \}.
\end{equation}

\begin{figure*}[ht]
    \centering
    \includegraphics[width=17cm,height=3.5cm]{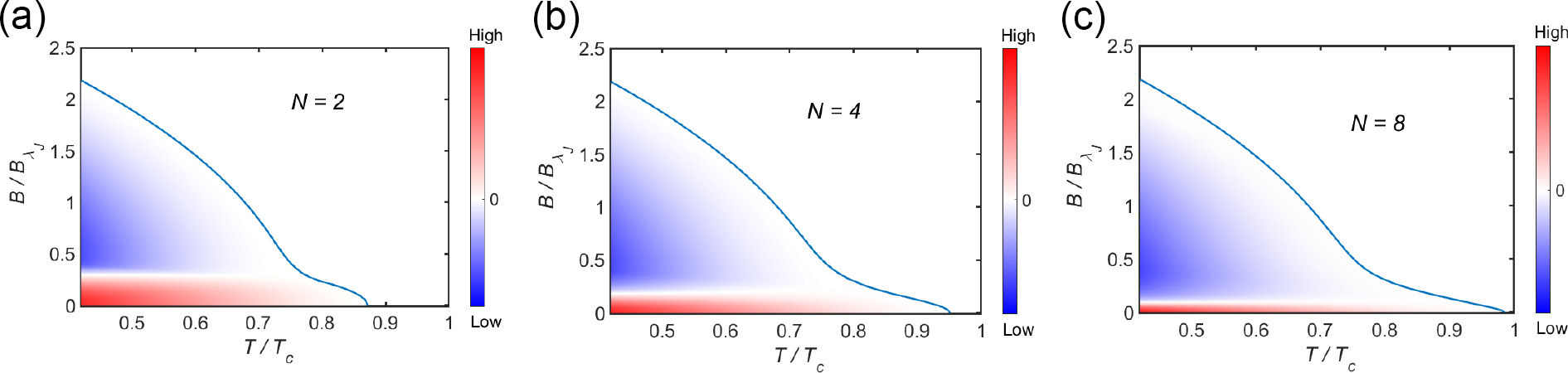}
    \caption{The distribution of the A state and B state in the superconducting phase for the layer numbers  $N=2$ (a), $N=4$ (b) and $N=8$ (c). The blue lines represents the upper critical field for each which are calculated according to \cite{Yuan2023}. The blue and red regions indicate the A state and B state, respectively, while the relative values represent the energy of A state minus that of the B state.}
    \label{figure S1}
\end{figure*}

Here \(\Bar{l} = l - (N+1)/2\) adjusts for the layer index relative to the center of the system. To optimize the free energy, we calculate the sets \(\{f_{l}(x)\}\) for the A state and \(\{f_{l}, q\}\) for the B state, under specified temperature and magnetic field conditions. We then compare their energies within the phase diagram to understand their relative stability. From Figs. \ref{figure S1}(a)-(c), we observe that the B state tends to occupy low magnetic field regions within the superconducting phase in finite systems. As anticipated, the A state becomes energetically favorable under strong magnetic field. Moreover, as the number of layers increases, the region occupied by the A state expands. The trend shown in Figs. \ref{figure S1} clearly indicates that if the number of layers approaches infinity, the A state could potentially occupy the entire superconducting regions.

\section{phase fluctuation of the dense Josephson vortex lattice}
In the main text, we derived the expression for \(P_l\) as shown in Equation (\ref{eq11}), assuming its small magnitude due to the dense configuration of the Josephson vortices. We also established a criterion for the validity of this expression. This section aims to demonstrate that \(P_l\) can closely approximate accurate numerical results when the ratio \(L_0^2/\lambda_J^2\) is moderate, which requires a relatively strong magnetic field. In Equation (\ref{eq9}), the periodicity in the out-of-layer direction results in \(P_{2n-2,2n-1} = P_{2n,2n+1}\), allowing us to deduce that \(P_{2n-1} = -P_{2n}\) and \(P_{2n-1,2n} = 2P_{2n} = -P_{2n,2n+1}\). Consequently, the second equation in Equation (\ref{eq9}) simplifies to

\begin{equation}\label{eqS21}
 \frac{\partial^{2} P_{2n}}{dx^2}+ \frac{1}{\lambda_{J}^{2}}   \left [  sin \left( \frac{2\pi x}{L_{0}} + 2P_{2n} \right)  - sin \left( \frac{2\pi x}{L_{0}} - 2P_{2n} +\pi \right) \right ] =0 .
\end{equation}
Equation (\ref{eqS21}) can be simplified and made dimensionless by introducing \(\Bar{x} = x/L_{0}\).

\begin{equation}\label{eqS22}
 \frac{\partial^{2} P_{2n}}{d\Bar{x}^2}+ \frac{2L_{0}^{2}}{\lambda_{J}^{2}}   sin \left( 2\pi \Bar{x} \right) cos \left( 2P_{2n} \right ) =0 .
\end{equation}
We investigate specific ratios of \(L_{0}/\lambda_{J}=1, 1.5, 2\), corresponding to \(B/B_{\lambda_{J}}=1, 0.67, 0.5\). The latter two values correspond to the blue and green points below the melting line in Figure \ref{figure 4}(c), respectively. Our analysis reveals minimal discrepancies between the accurate numerical results and the approximate analytical results, thus confirming the validity of our analytical expression.

\begin{figure*}[ht]
    \centering
    \includegraphics[width=17cm,height=4.5cm]{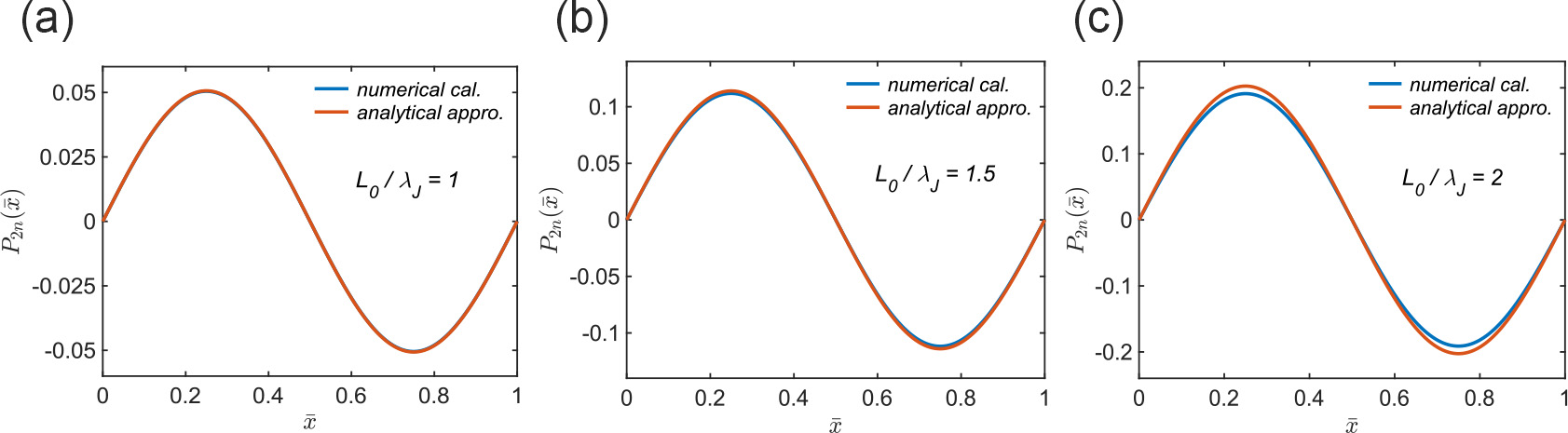}
    \caption{Comparison of numerical and approximate analytical results for the oscillation term \(P_{2n}\): (a) \(L_{0}/\lambda_{J}=1\), (b) \(L_{0}/\lambda_{J}=1.5\), (c) \(L_{0}/\lambda_{J}=2\).}
    \label{figure S2}
\end{figure*}

\end{document}